\documentclass[11pt,a4paper]{amsart}
\PassOptionsToPackage{table}{xcolor}
\usepackage[a4paper,margin=3cm]{geometry}
\usepackage{amsmath}
\usepackage{array}
\usepackage{graphicx}
\usepackage{longtable}
\usepackage{booktabs}
\usepackage{forest}
\usepackage{url}
\usepackage[authoryear, round]{natbib}
\usepackage[table]{xcolor}

\begin{document}

\title[Open-Source Software for Response-Adaptive Randomization]{Ready for the Clinic? A Survey of Open-Source Software for Response-Adaptive Randomization in Clinical Trials}
\author{Stina Zetterstrom\textsuperscript{1}, David S. Robertson\textsuperscript{1}, and Sof\'ia S. Villar\textsuperscript{1}}
\address{\textsuperscript{1} MRC Biostatistics Unit, University of Cambridge, United Kingdom}
	\email{stina.zetterstrom@gmail.com}
\date{\today}

\begin{abstract}

Response-adaptive randomization (RAR) modifies treatment allocation probabilities during a clinical trial as response/outcome data accumulate, with the aim of improving patient benefit, statistical efficiency, or both. Despite substantial methodological development, adoption of RAR in clinical practice has remained limited, and the software available to support its design and implementation has not previously been reviewed. We identified 16 publicly available, open-source software packages implementing RAR methods, spanning urn-based, target-allocation, Bayesian, Markov decision process (MDP)-based, and dose-finding approaches, and evaluated them with respect to their methodological and practical characteristics. We found that while several software packages exist, most are method-specific, and only a small number provide broader, general-purpose adaptive-trial design and analysis capabilities. Explicit support for practically relevant features, including delayed and missing outcome data, temporal trends in response rates, flexible operating-characteristic evaluation, and platform or multi-arm multi-stage trial designs, is rare or absent across the identified software. These findings suggest that while a diverse set of tools exists for exploring RAR designs, gaps remain between the methodological literature and the software available to implement it in practice. Continued development of flexible, practically oriented, and validated software is important for the wider adoption of RAR in clinical research, and we highlight interesting areas of further work.

\end{abstract}

\maketitle


\newpage

\section{Introduction}

Response-adaptive randomization (RAR) refers to a class of adaptive allocation procedures where treatment assignment probabilities are updated during the course of a trial based on accumulating response/outcome data. By preferentially allocating patients to treatments that appear more effective, RAR designs aim to improve patient benefit within the trial while maintaining the ability to draw valid statistical conclusions. Alternatively, depending on the design objectives, RAR can also prioritize statistical power or efficiency, and different procedures balance these objectives in different ways. For a recent overview of RAR methodology, see \citet{robertson2023response}.

Although methodological development in this area has been substantial, and interest in adaptive clinical trial designs grows, RAR has seen only limited use in practice \citep{robertson2023response}. Reasons for this may include the practical complexity of implementing a RAR allocation algorithm. This is a requirement specific to RAR, unlike other adaptive designs (e.g., incorporating arm-dropping or sample-size re-estimation in multi-arm multi-stage (MAMS) designs) that can be used with standard fixed randomization. While recent reviews have examined the use of RAR in clinical practice \citep{wilson2025response,sverdlov2025extent} and its methodological foundations \citep{robertson2023response}, no review has focused on the software available to implement these methods. As a result, it is unclear which methods are supported by existing software, how flexible current implementations are, and to what extent available tools address practical challenges of adaptive clinical trials.

The objective of this work is therefore to provide a review of existing open source software for RAR. Specifically, we identify available software, classify them according to their methodological scope and intended purpose, and evaluate the extent to which existing tools support important practical aspects of adaptive trial design and analysis. The software that we include in this review are: \texttt{adaptr} \citep{granholm2022adaptr, granholm2025designing}, \texttt{BAR} \citep{hsu2022bar}, \texttt{BATSS} \citep{couturier2026batss}, \texttt{BayesAdaptive} \citep{ventz2017bayesadaptive}, \texttt{blockRAR} \citep{chandereng2019robust}, \texttt{brar} \citep{pawel2025stabilizing}, \texttt{dfmta} \citep{riviere2026dfmtapkg}, \texttt{grouprar} \citep{zhai2024grouprar}, \texttt{MedianaDesigner} \citep{dmitrienko2023medianadesigner}, \texttt{RABR} \citep{zhan2022rabr}, \texttt{RARfreq} \citep{yu2024rarfreq}, \texttt{RARtool} \citep{ryeznik2015rartool}, \texttt{RARtrials} \citep{xu2025rartrials}, \texttt{SEARS} \citep{hsu2023sears}, \texttt{TrialMDP} \citep{merrell2021trialmdp}, and \texttt{TrialSimulator} \citep{zhang2026trialsimulator}.

The remainder of the article is organized as follows. Section~\ref{sec:RARoverview} provides a brief overview of RAR methods and reviews how RAR has been used in clinical trials to date. Section~\ref{sec:methods} describes the identification of software and the evaluation framework. The results of the software evaluation are presented in Section~\ref{sec:results}, followed by a discussion of key findings and future directions in Section~\ref{sec:discussion}.

\section{Overview of response-adaptive randomization} \label{sec:RARoverview}

This section provides a brief overview of the classes of RAR methods represented in the identified software. The aim is not to provide a comprehensive methodological review, but to introduce the main categories of methods required to understand the software.

\subsection{General framework} \label{sec:framework}

Unlike conventional fixed-randomization designs, where treatment assignment probabilities remain constant throughout the study, RAR designs update these probabilities as outcome data accumulate. This can improve patient benefit within the trial, statistical power to detect a treatment effect, or both, though these two goals often need to be balanced against each other \citep{rosenberger2015randomization}. This trade-off is often described as a balance between learning about treatment effects and assigning more patients to superior treatments. Different RAR procedures achieve this balance in different ways, resulting in a diverse methodological literature. A comprehensive overview of RAR methodology is provided by \citet{robertson2023response}.

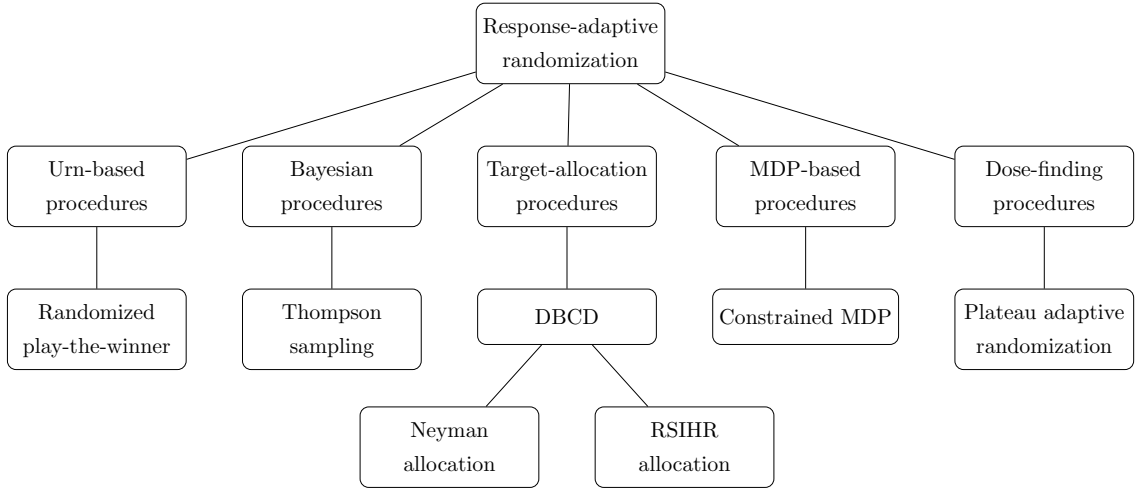
\begin{figure}[htbp]
\centering
\resizebox{\linewidth}{!}{%
\begin{forest}
for tree={
    draw,
    rounded corners,
    align=center,
    font=\small,
    minimum width=2.9cm,
    minimum height=0.9cm,
    inner sep=3pt,
    s sep=9mm,
    l sep=10mm,
    edge={-}
}
[Response-adaptive\\randomization
    [Urn-based\\procedures
        [Randomized\\play-the-winner]
    ]
    [Bayesian\\procedures
        [Thompson\\sampling]
    ]
    [Target-allocation\\procedures
        [DBCD
            [Neyman\\allocation]
            [RSIHR\\allocation]
        ]
    ]
    [MDP-based\\procedures
        [Constrained MDP]
    ]
    [Dose-finding\\procedures
        [Plateau adaptive\\randomization]
    ]
]
\end{forest}
}%
\caption{Simplified taxonomy of response-adaptive randomization (RAR) procedures. Boxes show representative examples of each class rather than an exhaustive list. See \citet{robertson2023response} for a comprehensive taxonomy.}
\label{fig:RAR}
\end{figure}

RAR procedures can be categorized into several methodological classes, including urn-based, target-allocation, Bayesian RAR, Markov decision process (MDP)-based, and dose-finding procedures. Figure~\ref{fig:RAR} illustrates the main classes of RAR procedures considered here. In short, urn-based procedures update treatment allocation probabilities according to predefined reinforcement rules and include classical randomized play-the-winner designs \citep{wei1978playthewinner}. Target-allocation procedures aim to converge toward a prespecified optimal allocation ratio, often through procedures such as the doubly adaptive biased coin design (DBCD), with common targets including the Neyman and RSIHR allocations \citep{rosenberger2015randomization}. Bayesian approaches use posterior distributions of treatment effects to guide allocation decisions, often through variants of Thompson sampling \citep{thompson1933likelihood}. MDP-based procedures, sometimes called bandit-based procedures, formulate the remaining trial as a sequential decision problem solved via dynamic programming \citep{baas2024constrained}. Dose-finding procedures, used in early-phase trials, aim to allocate patients to doses estimated to have the most favorable joint toxicity-efficacy trade-off. In addition to these, several extensions have been proposed to address practical considerations such as grouped patient recruitment, block randomization, and delayed outcome assessment. While the statistical methodology is mature, the implementation in trials depends in part on the availability of appropriate software. Lack of software can create two bottlenecks: software is needed for design exploration, to answer the question of whether RAR should be used, and, once that decision is made, software is needed to implement the design. Section~\ref{sec:practice} examines how RAR has been used in trial practice, and later sections assess the extent to which available software addresses these bottlenecks.

\subsection{RAR in practice} \label{sec:practice}

A continuously updated list of trials that have used RAR in some capacity is available in \citet{pin2024github}. Based on the trials in that list, RAR has been used in trials spanning a wide range of sizes, from the 12-patient ECMO trial \citep{bartlett1985extracorporeal}, an early application of randomized play-the-winner that resulted in a highly imbalanced final allocation which led to controversy, to the ongoing ACT-GLOBAL platform trial with a planned enrollment of 20,000 patients \citep{actglobal2024}. The typical (median) trial in the list, however, enrolled around 300 patients across three to four arms. Most trials are phase II, which typically involve fewer patients and more arms than large confirmatory phase III trials, and about three-quarters used some form of Bayesian RAR, i.e., using posterior probabilities to update allocation probabilities. Many of these trials are multi-arm platform designs, such as REMAP-CAP \citep{angus2020remap}, which spans 56 arms across multiple treatment domains, and I-SPY2 \citep{barker2009ispy2}, in which treatment arms are added or dropped during the trial. Two other reviews of RAR trials in practice have been reported by \citet{wilson2025response} and \citet{sverdlov2025extent}. Their results are similar to the list in \citet{pin2024github}, and report that RAR is predominantly used in phase II trials and in trials with more than two treatment arms, and that Bayesian methods are the most common approach. The number of reported RAR trials is growing and software that can handle the complexities of real trials is needed for accurate and reliable trial designs.

\section{Methods} \label{sec:methods}

\subsection{Identification of software}

Candidate software packages were initially identified through discussions with researchers active in the field of RAR. Additional software was identified through searches of CRAN and GitHub, published methodological reviews, reference lists of relevant articles, and other web searches. AI-assisted search tools (including ChatGPT, Gemini, and Claude) were also used to identify potentially relevant software and publications. Any software identified through AI-assisted searches was manually verified using package documentation and published references before inclusion.

Packages are included if they provided functionality for RAR in clinical trials, are publicly available, and have sufficient documentation to evaluate their methodological implementation and practical characteristics. Both dedicated RAR software and broader adaptive trial software providing RAR functionality are considered. Software developed primarily for general multi-armed bandit problems, reinforcement learning, or other non-clinical applications is excluded unless it also supports conventional clinical trial design and inference. Commercial software requiring licenses, such as FACTS (Berry Consultants) and East (Cytel), is not included so that the results of this review can be evaluated by the readers.

Several packages identified through these searches were considered but not included in the review. For instance, one actively maintained package, \texttt{bayesCT}, supports Bayesian adaptive trial design but does not implement RAR. Several other candidates were personal or course-project repositories rather than general-purpose, maintained software, including the deprecated \texttt{rarsim}.

For each included package, information was collected from package documentation, software repositories, and accompanying methodological publications.

\subsection{Evaluation framework}

The identified software is evaluated using a set of criteria covering both descriptive and practical aspects of RAR software. First, basic information was recorded for each package, including software availability, implementation language, maintenance status, scope, and primary software or methodological reference. The information collected on all software is summarized in Table~\ref{tab:description}.

The software is compared with respect to methodological and practical characteristics, see Tables~\ref{tab:comparison2a} and \ref{tab:comparison2b}. These include the allocation procedures implemented, supported outcome types, accrual timing, simulation of missing outcomes and temporal trends, early stopping rules, the ability to simulate a non-adaptive (fixed-randomization) benchmark design, customization options, and intended purpose. The aim is to compare the underlying randomization methods, and also to assess how flexible the software is and which aspects of trial conduct can be represented. In addition, several features that can be important for the practical use of RAR in trials are examined across all identified software, see Table~\ref{tab:practice}. These include delayed-outcome adaptation, adaptive missing-data handling (i.e., whether the allocation or inference procedures adjust for missingness, which is different from simulating its occurrence in Table~\ref{tab:comparison2a}), readiness for use in clinical trials, flexible operating-characteristic evaluation, user extensibility of the software and its implemented functions, and support for platform or MAMS trial designs. The selected criteria were motivated by experience in implementing RAR in practice  \citep{das2025implementing,das2025stratosphere} rather than by theoretical properties of the underlying randomization procedures. Because the reviewed packages differ in supported outcome types, allocation targets, and intended trial phase (Table~\ref{tab:description}), a single example applicable across all software was not feasible, and comparisons are instead based on each package's documented methodological and practical characteristics.

When conducting this review, the authors identified a small number of implementation issues. In \texttt{BAR} and \texttt{RARtrials}, these issues were identified through use of software in related work, and in several other packages, through code review with the assistance of Claude (Anthropic). All affected maintainers have been notified. These issues, and their status as maintainers respond, are tracked in an online repository, which is intended to be updated and open to reports of further issues from readers.\footnote{\url{https://github.com/StinaZet/RAR-software-bugs}}

\section{Results} \label{sec:results}

We first summarize descriptive characteristics of the identified software in Section~\ref{sec:overview}, and then compare methodological and practical characteristics in Section~\ref{sec:comparison}, before evaluating support for several practically important features in Section~\ref{sec:gaps}.

\subsection{Overview of available software} \label{sec:overview}

A total of 16 software packages satisfying the inclusion criteria are identified and Table~\ref{tab:description} summarizes these software together with availability, maintenance status, downloads, scope, trial phase, and primary reference. The majority are implemented in \texttt{R} and are available at CRAN, GitHub, or both. One package, \texttt{RARtool}, is implemented in \texttt{MATLAB}. \texttt{MATLAB} is not open source, but \texttt{RARtool} is included here as \texttt{RARtool} can be run in the open source platform \texttt{Octave} with minor changes to correctly call the functions. Last-update dates range from 2016 to 2026 and downloads vary between 0 and almost 59,000. However, this number includes automated downloads from CRAN for checks and should not be taken as a direct measure of the number of users.

The scope of the software is indicated using three categories describing how general the software is. \emph{Method-specific implementations} provide one or a few related RAR procedures, without supporting general trial configurations. Software denoted \emph{Simulation-oriented framework} is a general-purpose tool supporting a broad range of RAR procedures and trial configurations, primarily for comparing operating characteristics across designs. \emph{Broader adaptive-trial frameworks} include RAR as one component within a wider adaptive-trial-design toolkit that also covers non-RAR settings. The identified software includes eleven method-specific implementations (\texttt{BAR}, \texttt{BayesAdaptive}, \texttt{blockRAR}, \texttt{brar}, \texttt{dfmta}, \texttt{grouprar}, \texttt{RABR}, \texttt{RARfreq}, \texttt{RARtool}, \texttt{SEARS}, and \texttt{TrialMDP}), one simulation-oriented framework (\texttt{RARtrials}), and four broader adaptive-trial frameworks (\texttt{adaptr}, \texttt{BATSS}, \texttt{MedianaDesigner}, and \texttt{TrialSimulator}). Two packages (\texttt{dfmta} and \texttt{SEARS}) are focused on early-phase and dose-finding trials whereas the rest of the software focuses on later-phase trials.

Most software has a dedicated publication, either a methodological publication describing the methods in the software or a publication describing the software itself. However, two packages (\texttt{RARfreq} and \texttt{RARtrials}) implement methods from multiple publications without a dedicated reference, and two packages (\texttt{TrialSimulator} and \texttt{MedianaDesigner}) currently have no corresponding methodological or software publication.

\begingroup
\footnotesize
\setlength{\tabcolsep}{3pt}
\renewcommand{\arraystretch}{0.95}
\setlength{\LTcapwidth}{\linewidth}
\begin{longtable}{>{\raggedright\arraybackslash}p{2.5cm} >{\raggedright\arraybackslash}p{1.8cm} >{\raggedright\arraybackslash}p{1.05cm} >{\raggedright\arraybackslash}p{1.6cm} >{\raggedright\arraybackslash}p{2.8cm} >{\raggedright\arraybackslash}p{0.85cm} >{\raggedright\arraybackslash}p{2.9cm}}
\caption{Overview of software for response-adaptive randomization. Primary reference refers either to a dedicated software publication or to the principal methodological publication associated with the implemented methods.} \label{tab:description} \\
\hline
Software & Availability/ Language & Last update & Downloads & Scope & Trial phase & Primary\newline reference \\
\hline
\endfirsthead
\multicolumn{7}{l}{\small\textit{Table \thetable{} continued from previous page}} \\
\hline
Software & Availability/ Language & Last update & Downloads & Scope & Trial phase & Primary\newline reference \\
\hline
\endhead
\hline
\multicolumn{7}{r}{\small\textit{continued on next page}} \\
\endfoot
\hline
\endlastfoot
\texttt{adaptr}\newline (v1.5.0) & CRAN/ GitHub, \texttt{R} & Mar 2026 & 21,337 & Broader adaptive design framework & Late & \citet{granholm2022adaptr, granholm2025designing} \\
\texttt{BAR}\newline (v0.1.1) & CRAN/ GitHub, \texttt{R} & Nov 2022 & 11,665 & Method-specific implementation & Late & \citet{xiao2017bayesian} \\
\texttt{BATSS}\newline (v1.2.1) & CRAN/ GitHub, \texttt{R} & Sep 2026 & 6,576 & Broader adaptive design framework & Late & \citet{couturier2024batss, couturier2026batss} \\
\texttt{BayesAdaptive}\newline (GitHub) & GitHub, \texttt{R} & Mar 2017 & 1$^{\dagger}$ & Method-specific implementation & Late & \citet{ventz2017bayesian} \\
\texttt{blockRAR}\newline (v1.0.3) & CRAN/ GitHub, \texttt{R} & Jun 2026 & 18,737 & Method-specific implementation & Late & \citet{chandereng2019robust} \\
\texttt{brar}\newline (v0.1) & CRAN/ GitHub, \texttt{R} & Mar 2026 & 1,365 & Method-specific implementation & Late & \citet{pawel2025stabilizing} \\
\texttt{dfmta}\newline (v1.7-8) & CRAN, \texttt{R} & Mar 2026 & 59,943 & Method-specific implementation & Early & \citet{riviere2016dfmta} \\
\texttt{grouprar}\newline (v0.1.0) & GitHub/ CRAN, \texttt{R} & Mar 2024 & 5,335 & Method-specific implementation & Late & \citet{zhai2024group} \\
\texttt{MedianaDesigner}\newline (v0.13) & CRAN, \texttt{R} & Aug 2023 & 20,307 & Broader adaptive design framework & Late & - $^{\S}$  \\
\texttt{RABR}\newline (v0.1.1) & GitHub/ CRAN, \texttt{R} & Aug 2022 & 16,013 & Method-specific implementation & Late & \citet{zhan2021practical} \\
\texttt{RARfreq}\newline (v0.1.5) & CRAN, \texttt{R} & May 2024 & 20,768 & Method-specific implementation & Late & - $^{*}$ \\
\texttt{RARtool}\newline (N/A) & \texttt{MATLAB/ Octave} & 2016 & N/A$^{\ddagger}$ & Method-specific implementation & Late & \citet{ryeznik2015rartool} \\
\texttt{RARtrials}\newline (v0.0.2) & CRAN, \texttt{R} & Apr 2025 & 4,387 & Simulation-oriented framework & Late & - $^{*}$ \\
\texttt{SEARS}\newline (v0.1.0) & CRAN, \texttt{R} & Jun 2023 & 6,228 & Method-specific implementation & Early & \citet{pan2014sears} \\
\texttt{TrialMDP}\newline (GitHub) & GitHub, \texttt{R}/\texttt{C++} & Sep 2022 & 0$^{\dagger}$ & Method-specific implementation & Late & \citet{merrell2023trialmdp} \\
\texttt{TrialSimulator}\newline (v1.20.1) & CRAN/ GitHub, \texttt{R} & Jun 2026 & 4,305 & Broader adaptive design framework & Late & - $^{\S}$ \\
\end{longtable}
\endgroup

\footnotesize
\noindent \textit{$^{*}$ The package implements methods from multiple sources without a single summarizing reference.}\\
\textit{$^{\dagger}$ Not on CRAN, the figure shown is GitHub star count, not a download count.}\\
\textit{$^{\ddagger}$ Distributed only via the JSS archive, and therefore, no download counter is available.}\\
\textit{$^{\S}$ No peer-reviewed methodological or software publication exists for this package's RAR functionality and the documentation is limited to the package's own vignette/help pages or non-peer-reviewed technical manuals.}\\
\textit{Download counts (\texttt{cranlogs}, obtained 2026-09-23) correspond to CRAN's download counter, which can be inflated by automated mirror and package-checks and should be taken as an indication only.}

\normalsize

\subsection{Methodological and practical characteristics} \label{sec:comparison}

Tables~\ref{tab:comparison2a} and \ref{tab:comparison2b} summarize methodological and practical characteristics of the identified software. With the exception of \texttt{brar} (which computes randomization probabilities for a single decision point) and \texttt{RARtrials} and \texttt{TrialMDP} (which require user-written replication, or in the case of \texttt{TrialMDP}, provide exact policy computation rather than a built-in simulation loop), all identified software is oriented primarily toward operating-characteristic evaluation through simulation studies.

The methodological allocation procedures supported by current software are diverse, including urn-based, target-allocation, Bayesian, MDP-based, and dose-finding procedures. Binary and Gaussian outcomes are supported by most packages, whereas support for survival outcomes (\texttt{RARtool} and \texttt{dfmta}) and user-defined outcome types (\texttt{adaptr}, \texttt{BATSS}, \texttt{TrialSimulator}) are less common. The ability to simulate missing outcomes is implemented in three packages (\texttt{grouprar}, \texttt{MedianaDesigner}, and \texttt{TrialSimulator}), while handling of temporal trends is restricted to a single software (\texttt{blockRAR}).

There is also variety in the practical characteristics of the software, see Table~\ref{tab:comparison2b}. Trial duration can be modeled in four of the software (\texttt{dfmta}, \texttt{RARtool}, \texttt{RARtrials}, and \texttt{TrialSimulator}). It is worth noting that some software modeled accrual timing internally, but used it only to determine if an outcome is available in time to update the RAR algorithm, and do not output this information to the user. Early stopping of the trial can be done in nine of the 16 packages. A non-adaptive benchmark design exists explicitly in five packages (\texttt{BATSS}, \texttt{grouprar}, \texttt{MedianaDesigner}, \texttt{RARtool}, and \texttt{TrialSimulator}), and can be mimicked in all but one of the remaining packages by tuning the parameters of a RAR design to emulate a fixed design; the exception is \texttt{dfmta}, which provides no route, direct or indirect, to a fixed-allocation comparator. Inclusion of fixed designs is important as the ICH E20 guideline requires adaptive designs to be compared to a non-adaptive baseline design \citep{ich2025e20}. Furthermore, customization capabilities vary across packages. In Table~\ref{tab:comparison2b}, customization is categorized as \emph{Moderate} in software allowing users to adjust a few built-in tuning parameters (e.g., allocation targets, cohort sizes, thresholds). The software is categorized as having \emph{Extensive} customization capabilities if it has many tunable configurations, but is still limited to predefined settings. The software is characterized as \emph{User-extensible}  if users can supply their own custom functions to control, for instance, allocation rules and outcome-generation mechanisms. Following these criteria, \texttt{adaptr}, \texttt{BATSS}, \texttt{RARfreq}, and \texttt{TrialSimulator} provide the greatest degree of customization.

\renewcommand{\thetable}{\arabic{table}a}
\begingroup
\footnotesize
\setlength{\tabcolsep}{3pt}
\renewcommand{\arraystretch}{0.95}
\setlength{\LTcapwidth}{\linewidth}
\begin{longtable}{>{\raggedright\arraybackslash}p{2.6cm} >{\raggedright\arraybackslash}p{3.9cm} >{\raggedright\arraybackslash}p{4.7cm} >{\raggedright\arraybackslash}p{1.4cm} >{\raggedright\arraybackslash}p{0.9cm}}
\caption{Methodological characteristics of software for RAR.} \label{tab:comparison2a} \\
\hline
Software & Allocation methods & Outcome types & Missing\newline outcomes & Trend \\
\hline
\endfirsthead
\multicolumn{5}{l}{\small\textit{Table \thetable{} continued from previous page}} \\
\hline
Software & Allocation methods & Outcome types & Missing\newline outcomes & Trend \\
\hline
\endhead
\hline
\multicolumn{5}{r}{\small\textit{continued on next page}} \\
\endfoot
\hline
\endlastfoot
\texttt{adaptr} & Bayesian & Multiple (user-extensible) & No & No \\
\texttt{BAR} & Bayesian & Binary & No & No \\
\texttt{BATSS} & Bayesian & Multiple (user-extensible) & No & No \\
\texttt{BayesAdaptive} & Bayesian & Binary & No & No \\
\texttt{blockRAR} & Bayesian + frequentist, blocked & Binary & No & Yes \\
\texttt{brar} & Bayesian & Binary, Gaussian & No & No \\
\texttt{dfmta} & Dose-finding & Binary (toxicity); binary or survival (efficacy) & No & No \\
\texttt{grouprar} & Urn-based + target-allocation, grouped & Binary, Gaussian & Yes & No \\
\texttt{MedianaDesigner} & Bayesian & Normal & Yes & No \\
\texttt{RABR} & Rank-based & Binary, Gaussian & No & No \\
\texttt{RARfreq} & Urn-based + target-allocation & Binary, Gaussian & No & No \\
\texttt{RARtool} & Target-allocation & Survival & No & No \\
\texttt{RARtrials} & Bayesian + urn-based + target-allocation & Binary, Gaussian & No & No \\
\texttt{SEARS} & Dose-finding + Bayesian & Binary (toxicity, efficacy) & No & No \\
\texttt{TrialMDP} & MDP-based, blocked & Binary & No & No \\
\texttt{TrialSimulator} & User-defined$^{e}$ & Multiple (user-extensible) & Yes & No \\
\end{longtable}
\endgroup

\addtocounter{table}{-1}
\renewcommand{\thetable}{\arabic{table}b}
\begingroup
\footnotesize
\setlength{\tabcolsep}{3pt}
\setlength{\LTpost}{0pt}
\renewcommand{\arraystretch}{0.95}
\setlength{\LTcapwidth}{\linewidth}
\begin{longtable}{>{\raggedright\arraybackslash}p{2.8cm} >{\raggedright\arraybackslash}p{1.6cm} >{\raggedright\arraybackslash}p{2.3cm} >{\raggedright\arraybackslash}p{2.1cm} >{\raggedright\arraybackslash}p{2.1cm} >{\raggedright\arraybackslash}p{2.8cm}}
\caption{Practical characteristics of software for RAR.} \label{tab:comparison2b} \\
\hline
Software & Trial duration & Early stopping & Non-adaptive benchmark & Customization & Purpose \\
\hline
\endfirsthead
\multicolumn{6}{l}{\small\textit{Table \thetable{} continued from previous page}} \\
\hline
Software & Trial duration & Early stopping & Non-adaptive benchmark & Customization & Purpose \\
\hline
\endhead
\hline
\multicolumn{6}{r}{\small\textit{continued on next page}} \\
\endfoot
\hline
\endlastfoot
\texttt{adaptr} & No & Yes (efficacy, futility, equivalence) & Dedicated & User-extensible & Operating-charac\-teristic simulation \\
\texttt{BAR} & No & No & Via tuning parameters & Moderate & Operating-charac\-teristic simulation \\
\texttt{BATSS} & No & Yes (efficacy, futility) & Dedicated & User-extensible & Operating-charac\-teristic simulation \\
\texttt{BayesAdaptive} & No$^{d}$ & Yes (efficacy, futility) & Via tuning parameters & Moderate & Operating-charac\-teristic simulation \\
\texttt{blockRAR} & No & Yes (efficacy, futility) & Via tuning parameters$^{i}$ & Moderate & Operating-charac\-teristic simulation \\
\texttt{brar} & No & No & Via tuning parameters & Moderate & Allocation-probability calculation$^{g}$ \\
\texttt{dfmta} & Yes & Yes (no eligible dose remains) & No & Moderate & Operating-charac\-teristic simulation \\
\texttt{grouprar} & No$^{d}$ & No$^{a}$ & Dedicated & Extensive & Operating-charac\-teristic simulation \\
\texttt{MedianaDesigner} & No$^{b}$ & No$^{b}$ & Dedicated & Moderate & Operating-charac\-teristic simulation \\
\texttt{RABR} & No & No & Via tuning parameters & Moderate & Operating-charac\-teristic simulation \\
\texttt{RARfreq} & No & Yes (per-arm sample-size cap) & Via tuning parameters & User-extensible & Operating-charac\-teristic simulation \\
\texttt{RARtool} & Yes$^{h}$ & No & Dedicated & Moderate & Operating-charac\-teristic simulation \\
\texttt{RARtrials} & Yes$^{c}$ & Yes (futility) & Via tuning parameters & Moderate & Single-trial simulation$^{f}$ \\
\texttt{SEARS} & No & Yes (safety, futility, sample-size cap) & Via tuning parameters & Moderate & Operating-charac\-teristic simulation \\
\texttt{TrialMDP} & No & No & Via tuning parameters & Moderate & Optimal design (policy) computation \\
\texttt{TrialSimulator} & Yes & Yes (efficacy; futility user-coded) & Dedicated & User-extensible & Operating-charac\-teristic simulation \\
\end{longtable}
\endgroup
\renewcommand{\thetable}{\arabic{table}}
\footnotesize
\noindent \textit{$^{a}$ A stopping-rule implementation exists but is not exported to any user-facing function.}\\
\textit{$^{b}$ Implemented via separate modules (\texttt{FutRule}/\texttt{ADSSMod} for stopping, \texttt{EventPred} for duration) rather than the RAR module itself.}\\
\textit{$^{c}$ Patient accrual (Poisson arrivals) and outcome-observation delay are simulated internally and returned in the output, allowing trial duration to be derived manually; there is no automatic duration summary and no calendar-time-based stopping rule.}\\
\textit{$^{d}$ Patient accrual and/or response-timing information is simulated internally to drive the allocation or outcome-maturity logic, but no duration or timing summary is returned as an output.}\\
\textit{$^{e}$ Built-in randomization is fixed/permuted-block (optionally stratified); response-adaptive allocation rules are implemented by the user via custom action-function hooks calling \texttt{update\_sample\_ratio()}.}\\
\textit{$^{f}$ Each function simulates a single trial; operating characteristics such as power are obtained by the user via their own replication loop (e.g.\ \texttt{lapply}), not a function built into the package.}\\
\textit{$^{g}$ Computes randomization probabilities for a single decision point from cumulative data; the package does not include a built-in replication loop for evaluating operating characteristics.}\\
\textit{$^{h}$ Patient arrival times and calendar event/censoring times are simulated internally and returned in the output when run via the scripting interface (\texttt{trsimulate.m}), allowing trial duration to be derived manually; there is no automatic duration summary and no calendar-time-based stopping rule.}\\
\textit{$^{i}$ Setting \texttt{block\_number = 1} is explicitly documented as producing a traditional, non-adaptive RCT design, but this is a documented special case of a general block-structure parameter rather than a dedicated fixed-design argument.}

\normalsize

\subsection{Ecosystem-level characteristics} \label{sec:gaps}

The review in Section~\ref{sec:comparison} shows that many tools for RAR design exist. However, some functionality is missing for these software to be flexible enough for general trial design and analysis purposes. Table~\ref{tab:practice} summarizes several practical capabilities that may be desirable for implementation of RAR in modern clinical trials.

Delayed-outcome adaptation is explicitly supported in seven packages (\texttt{adaptr}, \texttt{BayesAdaptive}, \texttt{dfmta}, \texttt{grouprar}, \texttt{RARtool}, \texttt{RARtrials}, and \texttt{TrialSimulator}). Explicit handling of temporal trends is limited to a single package, \texttt{blockRAR}, and none of the identified packages can handle missing data beyond simulating its occurrence.

User extensibility, allowing users to define custom allocation rules, outcome-generation mechanisms, or other important trial components, is supported in only four packages (\texttt{adaptr}, \texttt{BATSS}, \texttt{RARfreq}, and \texttt{TrialSimulator}). Support for platform or MAMS trial designs is limited to \texttt{adaptr}, \texttt{BATSS}, and \texttt{BayesAdaptive}. Inclusion of platform trial capabilities is important as these are increasingly common practical applications of RAR. A composite or joint endpoint is implemented only by the two dose-finding packages. In \texttt{SEARS} this is implemented via a weighted toxicity-penalized efficacy score, and in \texttt{dfmta}, via a related joint toxicity-and-efficacy admissibility rule for dose selection.

None of the identified software provides the necessities to enable use in a live clinical trial, such as randomization of patients. Furthermore, with the exception of \texttt{TrialSimulator}, which allows users to supply their own \texttt{R} functions for analysis, no other software supports flexible, user-defined operating-characteristic evaluation beyond predefined inferential settings. In other words, 15 of the 16 packages are oriented toward simulation and methodological evaluation using fixed, built-in analysis approaches. In summary, this review indicates that, while several software packages provide flexibility for investigating complex adaptive designs, key challenges, including functionality for delayed and missing outcome data, temporal drift, and support for platform trial designs, are insufficiently addressed.

\begingroup
\footnotesize
\setlength{\tabcolsep}{3pt}
\setlength{\LTcapwidth}{\linewidth}
\begin{longtable}{>{\raggedright\arraybackslash}p{2.65cm} >{\raggedright\arraybackslash}p{2.6cm} >{\raggedright\arraybackslash}p{8.95cm}}
\caption{Summary of important practical gaps in current RAR software implementations.} \label{tab:practice} \\
\hline
Criterion & Packages satisfying & Key limitation \\
\hline
\endfirsthead
\multicolumn{3}{l}{\small\textit{Table \thetable{} continued from previous page}} \\
\hline
Criterion & Packages satisfying & Key limitation \\
\hline
\endhead
\hline
\multicolumn{3}{r}{\small\textit{continued on next page}} \\
\endfoot
\hline
\endlastfoot
\arrayrulecolor{gray}
Explicit delayed-outcome handling &
7/16\newline (\texttt{adaptr}, \texttt{BayesAdaptive}, \texttt{dfmta}, \texttt{grouprar}, \texttt{RARtool}, \texttt{RARtrials}, \texttt{TrialSimulator}) &
Most software assumes that all observations are available before randomizing the next patient(s) and does not provide support for evaluating RAR under delayed outcomes. \\ \hline

Missing-data handling &
0/16 &
Support is limited to simulation of missing outcomes, which is included in \texttt{grouprar} \citep{zhai2024group}, \texttt{MedianaDesigner}, and \texttt{TrialSimulator}, each of which can simulate missing or dropped-out data but does not adjust for it in the underlying analysis. \\ \hline

Live-trial support &
0/16 &
Current software is primarily intended for simulation and methodological evaluation rather than for use in ongoing clinical trials. \texttt{MedianaDesigner} provides a Shiny-based user interface, but this is only intended to increase user-friendliness. \\ \hline
 
Flexible operating-characteristic evaluation &
1/16\newline (\texttt{TrialSimulator}) &
Most software evaluates operating characteristics using predefined inferential settings, with limited support for alternative or user-defined methods. However, in \texttt{TrialSimulator}, users input their own \texttt{R} functions to perform the analysis and it is therefore not limited to a fixed set of built-in methods. \\ \hline

User extensibility &
4/16\newline (\texttt{adaptr}, \texttt{BATSS}, \texttt{RARfreq}, \texttt{TrialSimulator}) &
Only a few software packages allow user-defined allocation rules, outcome-generation mechanisms, or other important trial components. \\ \hline

Platform/\discretionary{}{}{}MAMS/\discretionary{}{}{}basket trial support &
3/16\newline (\texttt{adaptr}, \texttt{BATSS}, \texttt{BayesAdaptive}) &
Support for platform, MAMS, and basket trial designs is only included in a few general adaptive-trial packages. \\ \hline

Composite/\discretionary{}{}{}joint endpoint support &
2/16\newline (\texttt{SEARS}, \texttt{dfmta}) &
Support for a composite or jointly evaluated endpoint, in which multiple outcome components (e.g., efficacy and toxicity) inform a single allocation or selection decision, is rare and, where present, is specific to early-phase dose-finding designs. \\ 
\end{longtable}
\endgroup
\normalsize

\section{Discussion} \label{sec:discussion}

This review identified 16 publicly available software packages supporting RAR. We observe that many software packages for RAR already exist, reflecting continuing effort from independent research groups to make their methodological development available to the community. The 16 identified packages, many of them actively maintained, cover a wide range of RAR methods, including Bayesian, urn-based, target-allocation, MDP-based, and dose-finding approaches. We note that most software is oriented toward evaluating a method's operating characteristics rather than supporting its use in a live trial. Many packages implement a specific randomization procedure and provide limited support for alternative methods or broader adaptive-trial design and analysis. Furthermore, important practical challenges remain insufficiently addressed. Support for delayed outcomes and missing data is rare or nonexistent, and no software supports flexible, user-defined operating-characteristic evaluation or provides the interfaces typically needed for use in live clinical trials. User extensibility is also limited to a small number of packages. These issues represent important areas for future methodological and open software development.

Several limitations of this work should be acknowledged. The review focuses on publicly available software and did not provide a detailed evaluation of commercial adaptive-design tools. Furthermore, the assessment is largely based on package documentation and associated publications rather than hands-on testing of each software, so capabilities that are undocumented may have been missed. Also, documented features that do not function precisely as described, may be misrepresented in this review. Many of the discussed packages are also actively maintained, and other packages are also expected to be developed, and as a result, future developments can alter some of the classifications presented here.

These findings point to where new development effort would be best directed. Additional method-specific implementations of well-represented procedures, e.g., Bayesian or urn-based allocation for binary or Gaussian outcomes, offer comparatively little marginal benefit given how many such tools already exist. On the other hand, tools targeting the practical capabilities highlighted as missing above are much needed. This includes capabilities for delayed and missing outcome data, temporal drift, flexible operating-characteristic evaluation, and platform or MAMS trial support. Broader adaptive-trial software such as \texttt{adaptr}, \texttt{BATSS}, \texttt{TrialSimulator}, and \texttt{MedianaDesigner} may indicate a shift toward more flexible software of this kind. As a final remark, we note that a wide range of software for RAR is already available, supporting many methodological approaches and simulation-based evaluations, and building on it to close the remaining gaps will be important for the broader adoption of RAR in clinical research.

\section*{Acknowledgments}

We are grateful to the developers of the software packages reviewed in this work. The breadth and quality of what is already publicly available made this review possible, and we thank them for the sustained effort of building and maintaining these tools openly for the research community.

\section*{Competing interests}

The authors are currently developing an open-source package for response-adaptive randomization (\texttt{RARchitect}, in early development and not yet publicly released), and this review was conducted in part as background work for that development. The \texttt{BAR} and \texttt{RARtrials} issues arose from separate, unrelated work and are not connected to \texttt{RARchitect}. The authors are colleagues with a developer of \texttt{BATSS}, but this relationship did not influence its evaluation. The authors declare no other competing interests.

\section*{Funding}

This work was supported by the Medical Research Council, grant number MR/Z503538/1.

\section*{Data Availability}

This review did not generate new data or software and no new data or code are associated with this manuscript.

\bibliographystyle{plainnat}
\bibliography{ref}

\end{document}